\documentclass[prd,aps,twocolumn,lengthcheck,showpacs,preprintnumbers,
nofootinbib,letterpaper,amsmath,amssymb,superscriptaddress,nofootinbib]%
{revtex4-1}

\usepackage{amsbsy}
\usepackage[nolist,nohyperlinks]{acronym}
\usepackage{latexsym}
\usepackage{graphicx}
\usepackage{url}
\usepackage{xcolor}

\begin{document}

\title[Localization with future GW networks]
{Localization of binary
mergers with gravitational-wave detectors of second and third generation}

\author{Joseph Mills, Vaibhav Tiwari and Stephen Fairhurst}

\address{Cardiff School of Physics and Astronomy,
Cardiff University, Queens Buildings, The Parade, Cardiff CF24 3AA, UK
}

\begin{abstract} 
The observation of gravitational wave signals from binary black hole mergers has established the field of 
gravitational wave astronomy. It is expected that future networks of gravitational wave detectors will possess 
great potential in probing various aspects of astronomy. An important consideration for successive improvement 
of current detectors or establishment on new sites is knowledge of the minimum number of detectors 
required to perform precision astronomy. We attempt to answer this question by assessing ability of future detector 
networks in detecting and localizing binary neutron stars mergers in the sky. This is an important aspect as a good localization ability is crucial for many of the scientific goals of gravitational wave astronomy, such as electromagnetic follow-up, measuring the properties of compact binaries throughout cosmic history, and cosmology. We find that although two 
detectors at improved sensitivity are sufficient to get a substantial increase in the number of observed 
signals, at least three detectors of comparable sensitivity are required to localize majority of the signals,  
typically to within around  $10 \deg^{2}$ --- adequate for follow-up with most wide field of view optical 
telescopes.

\end{abstract}

\maketitle

\acrodef{GW}{gravitational wave}
\acrodef{BH}{black hole}
\acrodef{BBH}{binary black hole}
\acrodef{BNS}{binary neutron star}
\acrodef{NSBH}{neutron star--black hole}
\acrodef{EM}{electromagnetic}
\acrodef{NS}{neutron star}
\acrodef{FoV}{field of view}
\acrodef{PCA}{principal component analysis}
\acrodef{PN}{post-Newtonian}
\acrodef{SNR}{signal-to-noise ratio}
\acrodef{PE}{parameter estimation}
\acrodef{MCMC}{Markov chain Monte Carlo}
\acrodef{PSD}{Power Spectral Density}
\acrodef{GRB}{Gamma Ray Burst}

\newcommand{\PN}[0]{\ac{PN}\xspace}
\newcommand{\EM}[0]{\ac{EM}\xspace\xspace}
\newcommand{\BH}[0]{\ac{BH}\xspace}
\newcommand{\NS}[0]{\ac{NS}\xspace}
\newcommand{\PCA}[0]{\ac{PCA}\xspace}
\newcommand{\GW}[0]{\ac{GW}\xspace}
\newcommand{\SNR}[0]{\ac{SNR}\xspace}
\newcommand{\PE}[0]{\ac{PE}\xspace}
\newcommand{\MCMC}[0]{\ac{MCMC}\xspace}

\section{Introduction}
\label{sec:intro}

One hundred years after gravitational waves were predicted \cite{Einstein:1918}, the first detection of a
\ac{BBH} coalescence by the advanced LIGO detectors \cite{Abbott:2016blz} has heralded the beginning of the era of 
\ac{GW} astronomy.  Over the coming years, the sensitivity of the advanced LIGO instruments will improve
and the advanced Virgo, KAGRA and LIGO India detectors will join the global network \cite{collaboration_advanced_2015, Acernese:2015aa, Aasi:2013wya, Reitze:2011, 
aso:2013eba}.  This network of
advanced gravitational wave detectors is expected to observe many more \ac{BBH} mergers, 
as well as \ac{GW} emitted during the merger of \ac{BNS} and \ac{NSBH} binaries
 \cite{TheLIGOScientific:2016pea, Abbott:2016ymx}.  Additionally, \ac{GW} emitted by
non-symmetric neutron stars, core-collapse supernovae, and other astrophysical transient events may be observed
\cite{Abbott:2016ezn, Abbott:2016cjt}.

A major goal of \ac{GW} astronomy is the observation of an \ac{EM} counterpart to a signal.  Indeed, although
\ac{BBH} mergers are not expected to produce an electromagnetic signal, a broadband follow-up of GW150914 
demonstrated the willingness of the wider astronomical community to engage in multi-messenger observation of GW 
sources \cite{Loc_and_broadband}. Unlike most EM telescopes, GW detectors are not pointing instruments, and 
localization is achieved primarily by measuring the differences in arrival times of the signal in different detectors 
\cite{Fairhurst:2009tc}. Consequently, searching the relatively large GW localization regions (630 $\deg^2$ for 
GW150914 \cite{Loc_and_broadband}) represents a challenge for even wide \ac{FoV} telescopes.
These telescopes have \acp{FoV} on the order of $10 \deg^2$ or less \cite{Rau:2009yx, panstarrs, skymapper}. The 
ability to confidently identify \ac{EM} counterparts to \ac{GW} events will depend on the GW localization, which should be 
comparable to the telescope FoV.  The addition of Virgo, and further detectors in India and Japan, improves the 
localization ability of the network greatly, allowing many signals to be localized to within tens of square degrees
\cite{Aasi:2013wya, fairhurst:2014, Singer:2014qca}.

Compact binary systems composed of at least one neutron star have plausible EM counterparts across gamma, 
x-ray, optical, infrared, and radio bandwidths (for possible counterparts see 
\cite{chassande-mottin_multimessenger_2011, cowperthwaite_comprehensive_2015, kanner_LOOC_2008, 
lee_gravitational_2015, Metzger:2011bv, nissanke_identifying_2013, nuttall_identifying_2010, stubbs_linking_2008, sylvestre_prospects_2003}). 
Synergistic GW and EM astronomy will increase confidence in associations between \ac{GW} and \ac{EM} signals,
and provide complementary information. For instance, observing a near-infrared kilonova associated with a 
\ac{BNS} would firmly establish the progenitor \cite{Tanvir:2013pia, Berger:2013wna}.  Furthermore, it would enable the 
determination of the host galaxy
which, in turn, allows for independent measurements of the luminosity distance and redshift of the source \cite{Schutz:1986gp}.  
This would allow precision tests of cosmology and the variation of the dark energy equation of state with redshift \cite{sathyaprakash_cosmography_2010}.

There are plans for future gravitational wave detectors that will be significantly more sensitive than the current
generation of advanced detectors.  These include upgrades to the existing detectors, such as $A+$ and LIGO 
Voyager \cite{ISwhitepaper:2016}, which gives the best possible sensitivity within the current LIGO infrastructure.
Additionally, entirely new detectors have been proposed.  The Einstein Telescope is a next-generation
European gravitational wave observatory \cite{punturo:2010zz, punturo:2010zza, punturo:2010zzb}, and 
Cosmic Explorer \cite{Evans:2016mbw} is a proposed US-based future detector, both of which improve on
the advanced detector sensitivity by a factor of ten or more.  
As well as revealing new sources of gravitational waves, these detectors
will allow us to observe \ac{BBH} mergers throughout most of the history of the universe and \ac{BNS} to
cosmological distances.  Furthermore, the nearby signals will be very loud in these detectors, allowing
for unprecedented tests of Einstein's general relativity, and observation of matter at supra-nuclear density
inside neutron stars.  As with the advanced detector network, joint \ac{GW}-\ac{EM} observations will be vital
in fully extracting the science from these observations \cite{fan_multimessenger_2015, nissanke_identifying_2013, cowperthwaite_comprehensive_2015, Phinney:2009ty}.

The science case for these new facilities is still evolving, and will continue to do so as further gravitational
wave observations are made.  Estimates of the accuracy with which networks of third and second generation 
detectors can reconstruct parameters will inform decisions over the viability of new facilities. There have been 
previous studies of ET that estimate the detection efficiency and the accuracy of mass measurements
\cite{regimbau_revisiting_2015, regimbau_second_2014, regimbau_mock_2012, meacher_second_2016}. 
Estimates of the localization ability of various third generation networks were also considered as part of a 
comprehensive parameter estimation study \cite{Vitale:2016icu}.  Furthermore, detailed studies of the optimal
location of future detectors have been performed \cite{raffai_optimal_2013, Blair:2008zz, hu_global_2015}.

One practical consideration is whether it would be advantageous to accelerate the development of third 
generation detectors, perhaps at the expense of further upgrades to the second generation, or if the
operation of a heterogeneous network of detectors is preferable.  To date, there is rather little in the literature
on the merits of such networks.  Here we investigate the differences between homogeneous and heterogeneous
networks of detectors.  For concreteness, we focus primarily on the sky coverage of the networks and the 
accuracy with which they are able to localize sources.  We consider the network localization accuracy for 
both face-on \ac{BNS} systems at a fixed distance as well as a population of \ac{BNS} distributed 
isotropically and uniformly in comoving volume.

Previous estimates of network localization errors largely fall into two distinct categories: the first being analytical 
estimates that bypass the full task of parameter estimation and reduce the parameter 
space by focusing primarily on source localization \cite{Fairhurst:2009tc, Blair:2008zz, cavalier_reconstruction_2006, 
PaiDhurandharBose2001, 
sylvestre_position_2004, klimenko_localization_2011, schutz_networks_2011, wen_geometrical_2010, 
fairhurst_source_2011, Singer:2014qca, fairhurst:2014, sathyaprakash_ligo-t1000251-v1:_2010, iyer_ligo-india:_2013}; 
the second being full parameter estimation studies that extract detailed parameter estimates using Bayesian 
statistics \cite{rover_coherent_2007, van_der_sluys_parameter_2009, raymond_degeneracies_2009, 
nissanke_localizing_2011, veitch_estimating_2012, veitch_parameter_2015,  
berry_parameter_2015, rodriguez_basic_2014, singer_rapid_2016}. Performing the full analysis has the advantage
of being more accurate, but due to the computational cost the number of sources that can be considered is typically
small.  On the other hand, 
analytical studies using only the timing information \cite{Fairhurst:2009tc, fairhurst_source_2011} have been shown to 
overestimate the localization error region \cite{grover_comparison_2014}. Here, we make use of an improved, analytical 
method that incorporates amplitude and phase consistency between the sites, as well as timing \cite{fairhurst:2017}.  

This paper is organized as follows. Section \ref{sec:networks} will describe the networks used in this study. Section 
\ref{sec:loc} introduces the method for calculating the localization error regions. We present and analyse our main 
results in Section \ref{sec:results} before concluding in Section \ref{sec:discussion}.

\section{Future Detectors and Networks}
\label{sec:networks}

\subsection{Future detectors}

\ac{GW} detector sensitivity is limited by a number of fundamental noise sources. These noise sources can be broadly separated into 
two categories: displacement noise and sensing noise. Displacement noises cause motions of the test masses. 
Noise sources such as seismic noise and mechanical resonances come in this category. Sensing noises,
on the other hand, are phenomena that limit the ability to measure those motions; they are present even in the absence of test mass 
motion. Shot noise and thermal noise are included in this category.  In addition, there are technical noise sources which must be understood 
and mitigated in order that the detector sensitivity is limited by fundamental noise.  Typically, low frequency sensitivity is limited by 
seismic noise, mid frequencies are limited by thermal noise and higher frequencies are limited by quantum noise. LIGO underwent a 
series of upgrades from its initial to advanced configuration to address each of the noise sources \cite{Aligo:2014}. 
Seismic noise is being suppressed by the use of multi-stage mechanical seismic isolation and quadruple pendulum suspension systems. 
Thermal noise arises in test masses and suspensions and is determined by material properties and beam size. Compared to initial LIGO, 
advanced LIGO uses a  larger beam size. This results in better averaging of beam on a larger surface area which combined with better 
coating and suspension material results in efficient dissipation of heat.  Quantum noise arises due to statistical fluctuations in detected 
photon arrival rate. Quantum noise is overcome by increasing the beam power and increasing the weight of the test masses to overcome 
the increased radiation pressure.

Many technologies have been proposed to further increase the sensitivity of ground based detectors. For example, building detectors 
underground to suppress gravity gradients \cite{punturo:2010zz}, 
improving mirror coatings (Section 5.9.3 in \cite{ET}) and cryogenically cooling the mirrors for reducing the thermal, 
using squeezed light for lowering the noise floor due to quantum noise \cite{squeezing}.  A detailed discussion on possible technology 
improvements is given in \cite{ET}.
In the following, we briefly introduce several proposed future detector configurations and their corresponding sensitivities.  These are
used in the following sections when comparing the performance of different networks.

\textit{LIGO Voyager:} Various upgrades have been proposed for the advanced LIGO detectors \cite{voyager:2016} leading to the 
proposal for an upgrade to $A+$ in 2020 followed by a further upgrade to LIGO Voyager which is envisioned to be operational 
around 2025.  Voyager improves on the sensitivity of advanced LIGO by around a factor of three across a  broad frequency range. The increased sensitivity is intended to be achieved by improvements in all the departments (seismic isolation system, coatings of mirrors, heavier and larger test 
masses, increased beam power, etc.) of the advanced LIGO infrastructure combined by frequency dependent squeezing and cryogenic cooling of mirrors \cite{LIGO3Pendulum, LIGO3cryo, LIGO3susp}. 

\textit{Einstein Telescope:} Various studies have shown that further increase in sensitivity is required for performing precise gravitational wave astronomy, testing of general relativity and 
improving our understanding of exotic phenomenon like equation of state and tidal deformability of neutron stars \cite{chassande-mottin_multimessenger_2011, BS:2010, Hllr:2010, BS:2012, Cy:2017}. The Einstein 
Telescope is a proposed next-generation European gravitational wave observatory \cite{punturo:2010zz, punturo:2010zza, punturo:2010zzb} with sensitivity an order of magnitude higher than advanced 
LIGO and extending down to $1 \mathrm{Hz}$. It intends to achieve this improvement through a combination of longer arms and improved technologies. The  original 
design called for a triangular configuration of three interferometers with 10 km arms and $60^{\circ}$ angle between the arms. In addition, the proposed \textit{xylophone} configuration allows installation of separate 
high and low-frequency detectors. High frequency sensitivity is most easily achieved with high laser power, but this generates significant complications at lower frequencies. The divided detector avoids this issues 
by allowing to pursue different strategies in optimising the noise for each frequency range. Additionally, it also reduces the length of tunnel required (as each tunnel is used by two of the interferometers) and also 
makes the detector sensitive to both gravitational wave polarizations \cite{ET}.

\textit{Cosmic Explorer:} There is also a proposal for a Cosmic Explorer detector \cite{Evans:2016mbw, ISwhitepaper:2016}, which would be around a factor of three more sensitive than ET. The design and technology used 
is similar to ET but with arm length that can stretch out to between 40 to 50 km. Although the possibility of these detectors only lies in the far future, it is noteworthy that these detectors can see GW150914 like BBH 
mergers throughout the visible universe.

\begin{figure}[tb]
\centering
\includegraphics[width=1.0 \linewidth]{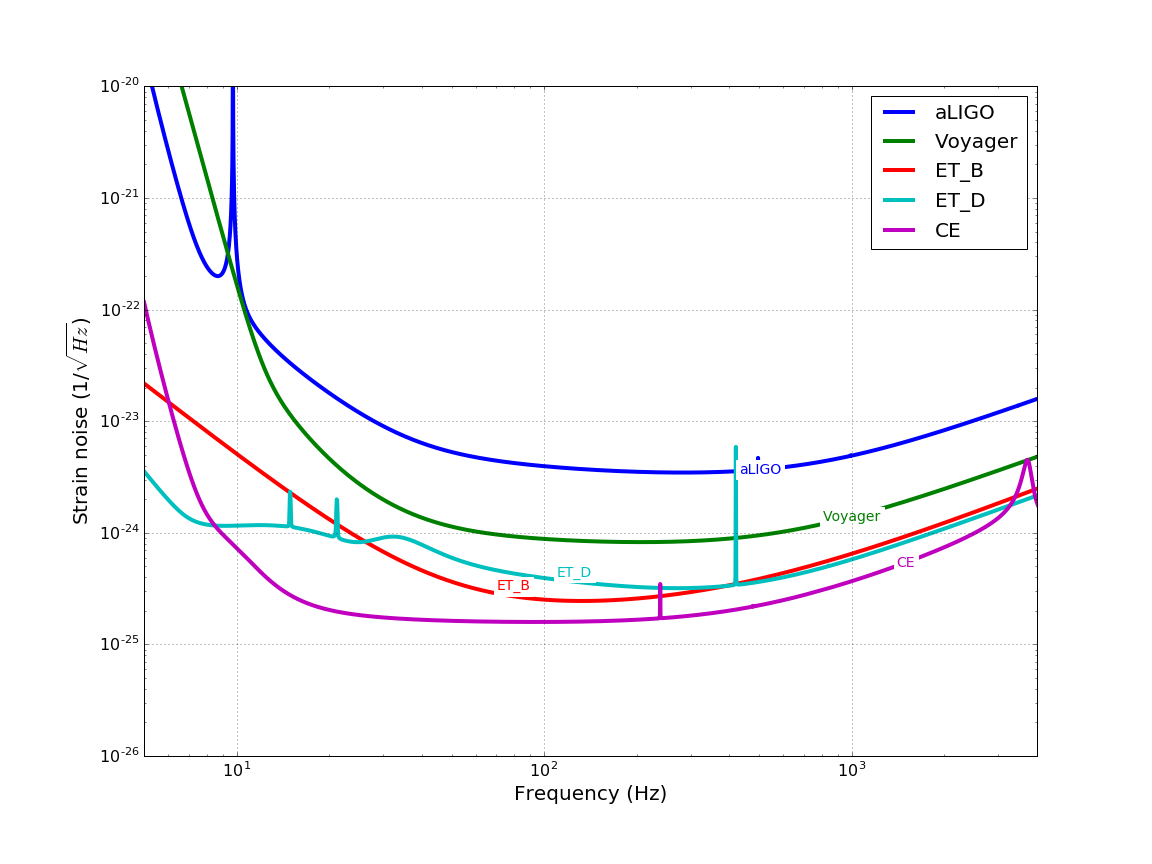}
\caption{Target noise curves for future detectors \cite{Evans:2016mbw}: 
advanced LIGO at design sensitivity (aLIGO); 
LIGO Voyager; Einstein Telescope (two proposed configurations, ET-B and ET-D) and Cosmic Explorer (CE).}
\label{fig:spectra}
\end{figure}

In figure \ref{fig:spectra} we show the sensitivities of the proposed future detectors \cite{Evans:2016mbw}, as well as the advanced LIGO
design sensitivity. We show the ET \textit{xylophone} configuration, called ET-D. Also included for comparison is ET-B, 
which is an alternative ET configuration where every interferometer is optimized for best overall sensitivity, 
but at the expense of some low frequency sensitivity. For all ET simulations in this study, ET-D sensitivity is assumed.  

\subsection{Networks}

We will consider five networks of gravitational wave observatories beyond the advanced detectors that
are currently being built, commissioned and operated.  Specifically, we consider:
\begin{enumerate}
\item[(i)] A network comprising detectors at the three LIGO sites (Hanford, Livingston and India) 
where the detectors have been upgraded to LIGO Voyager sensitivity. (Voyager)
\item[(ii)] A network comprising the three LIGO Voyager detectors complemented by a triangular ET detector
in Europe. (Voyager-ET)
\item[(iii)] A network with three L-shaped detectors at ET sensitivity distributed globally. (3ET)
\item[(iv)] A network comprising a triangular ET detector and two Cosmic Explorer detectors (CE-ET)
\item[(v)] A network of three Cosmic Explorer detectors (3CE).
\end{enumerate}

Networks (i) and (ii) arise naturally from existing proposals, but there is currently no global plan for a third generation
network. Although there is no proposal for a network
of ET detectors, we include this as configuration (iii), to facilitate comparison with the Voyager-ET network.
For simplicity, we use three L-shaped ET detectors, with a sensitivity matching
the proposed triangular ET detectors.%
\footnote{It can be shown that the triangular ET detector has the same sensitivity as two co-located 
L-shaped detectors of length 10.6 km whose orientations differ by $45^{\circ}. $\cite{punturo:2010zz}  
The network of L-shaped ET detectors then
provides three detectors of comparable sensitivity to the two \textit{effective} L-shaped detectors in the triangular ET.  
However, this leads to a doubling in the length of tunnels required.  In the case 
where the cost of constructing the tunnels is dominant, one could instead construct two 7km interferometer within the same tunnel
making use of each tunnel twice, as is done in the triangular ET design.  In this scenario the tunnel length is increased by around 50\%.}
We also consider a comparable network comprised of Cosmic Explorer detectors as well as a heterogeneous CE-ET network.  
Both the Voyager-ET and CE-ET network exhibits substantial heterogeneity of sensitivities 
with a factor of three difference over a broad frequency range.
The majority of previous studies, have assumed that the detectors in the network have identical sensitivity
\cite{fairhurst:2014, fairhurst_source_2011, Fairhurst:2009tc, raffai_optimal_2013}.  

The locations of future gravitational wave detectors have not yet been finalized.  In this study make use of
the detector locations derived in \cite{raffai_optimal_2013} to optimize the location of future detectors. 
There, a three part Figure of Merit is used to determine the optimal location of detectors in a network, 
comprising equal parts: 1) How equally the network can determine both polarizations; 
2) a simple measure of localization ability based on the area of the triangle formed by the telescopes and
3) how accurately the chirp mass can be measured.  The locations and orientations of all detectors are reported in Appendix
\ref{sec:appendix}.

For the LIGO Voyager network, the location of the Hanford and Livingston detectors is fixed.  Their orientations
were chosen so that they were, as much as possible, sensitive to the same gravitational wave polarization,
thus improving the chances of coincident detection \cite{Blair:2008zz}.  The location of the LIGO India
detector has not been announced at this time, so we use the optimal location from \cite{raffai_optimal_2013},
which places it in a seismically quiet location.
The triangular ET detector is added to this network to form the Voyager-ET network.  In \cite{raffai_optimal_2013}, it was shown that
a location in Slovakia gave maximum flexibility when constructing a global network, so we choose this.  Since
ET is equally sensitive to both gravitational wave polarizations, the orientation of the detectors does not affect the results.
It should be noted that the precise location in Europe of the triangular ET does not have a significant impact on results.

For the 3ET and CE networks, we are free to optimally site all three of the new detectors.  In \cite{raffai_optimal_2013}, 
with the additional requirements that the detectors lie on the land and avoid areas with a high degree of human activity,
the authors arrived at two comparable networks for three triangular ET detectors.  The best configurations
had detectors in either Australia, Central Africa and the USA or in Australia, Europe and South America.  Although the
optimization was performed for triangular ET detectors, we use the first set of locations for both the 3ET and CE networks.  
We then optimize the orientation of the detectors based on part 1 of the figure of merit --- sensitivity to both gravitational 
wave polarizations --- as parts 2 and 3 will be largely insensitive to the orientation.
Finally, for the CE-ET network, we retain the two CE detectors
in the USA and Australia and augment the network with a triangular ET detector in Europe.

This by no means covers the full set of proposed future detectors and networks, but is sufficient to allow us 
to explore the impact of a heterogeneous set of detector sensitivities and compare this to networks where all 
detectors have the same, or similar, sensitivity.  

\subsection{Network Sensitivity}

The response of a detector to the two polarizations of a gravitational wave is given by $F_{+}$ and $F_{\times}$,
which are functions of the sky location and polarization of the wave \cite{thorne.k:1987}.  For networks of equally sensitive detectors, the network
response at a given sky point is given by $\left[ \sum_{i}  (F_{+}^{i})^{2} + (F_{\times}^{i})^{2} \right]^{1/2}$ 
\cite{schutz_networks_2011, raffai_optimal_2013}.  However, when dealing with heterogeneous networks, we must generalize the expression to take account of the detector sensitivity.
To do so, we introduce a sensitivity measure $\rho_{o, i}$ defined as \cite{Allen:2005fk}%
\footnote{This quantity is often denoted $\sigma$.  To avoid confusion with the signal bandwidth, $\sigma_{f}$, 
introduced in the next section, we use the notation $\rho_{o}$ here.}
\begin{equation}
  \rho_{o, i}^{2} = 4 \int_{0}^{\infty} \frac{|\tilde{h}_{o} (f)|^{2}}{S_{i}(f)} df
\end{equation}
where $\tilde{h}_{o}(f)$ is the gravitational wave strain from a fiducial system placed overhead the detector at a fixed
distance and face-on, and $S_i(f)$ is the \ac{PSD} of the detector noise.  Then $\rho_{o, i}$ gives the expected SNR for 
such a signal in detector $i$. For our study, we take 
$\tilde{h}_{o}(f)$ to be the signal from a face-on binary neutron star system at 1 Mpc from the detector.  Then, we weight
the response of each detector by the sensitivity, defining \cite{harry_targeted_2011}
\begin{equation}
  f^{i}_{+, \times} = \rho_{o,i} F^{i}_{+, \times} \, .
\end{equation}
The relative sensitivity of the network at a given sky point is then defined as the network response,
\begin{equation}\label{eq:response}
N_{R}=\left( \frac{\sum_{i}  \left[ (f_{+}^{i})^{2} + (f_{\times}^{i})^{2} \right]  }{ \sum_{j}  \rho_{o, j}^{2}}  \right)^{1/2} \, ,
\end{equation}
where the indices $i,j$ run over the detectors.  Using this definition, the maximum network response is unity and this
will only be achieved when all detectors are aligned to be maximally sensitive to the same sky position.
This extends the definition of \cite{schutz_networks_2011} to
a heterogeneous network and is closely related to the network sensitivity to generic transients introduced in
\cite{klimenko_localization_2011}.  

We are also interested in the relative sensitivity to the two gravitational wave polarizations.  To define this
unambiguously, we must identify a preferred choice of the $+$ and $\times$ polarizations or, equivalently,
a choice of polarization angle.  We define the Dominant Polarization Frame \cite{klimenko_constraint_2005, 
harry_targeted_2011}, which gives the maximum sensitivity to the $+$ polarization.  To do so, we introduce
\begin{equation}
  f^{\mathrm{net}}_{+, \times} = \left( f^{1}_{+, \times}, \ldots, f^{N}_{+, \times} \right) \, .
\end{equation}
The dominant polarization frame, for a given sky location, is the unique frame such that: 
(1) $f_{\times}^{\mathrm{net}} \cdot f_{+}^{\mathrm{net}}=0$; (2) the network is maximally sensitive to the $+$
polarization, thus ensuring $|f_{+}^{\mathrm{net}}|\geqslant |f_{\times}^{\mathrm{net}}| $. The ratio of 
$|f_{\times}^{\mathrm{net}}| $ to $|f_{+}^{\mathrm{net}}|$ is called the network alignment factor 
\cite{klimenko_constraint_2005} and will vary from one --- equal sensitivity to both
polarizations --- to zero --- sensitivity to a single polarization.   

\begin{figure*}[tb]
\centering
\includegraphics[width=1.0 \textwidth]{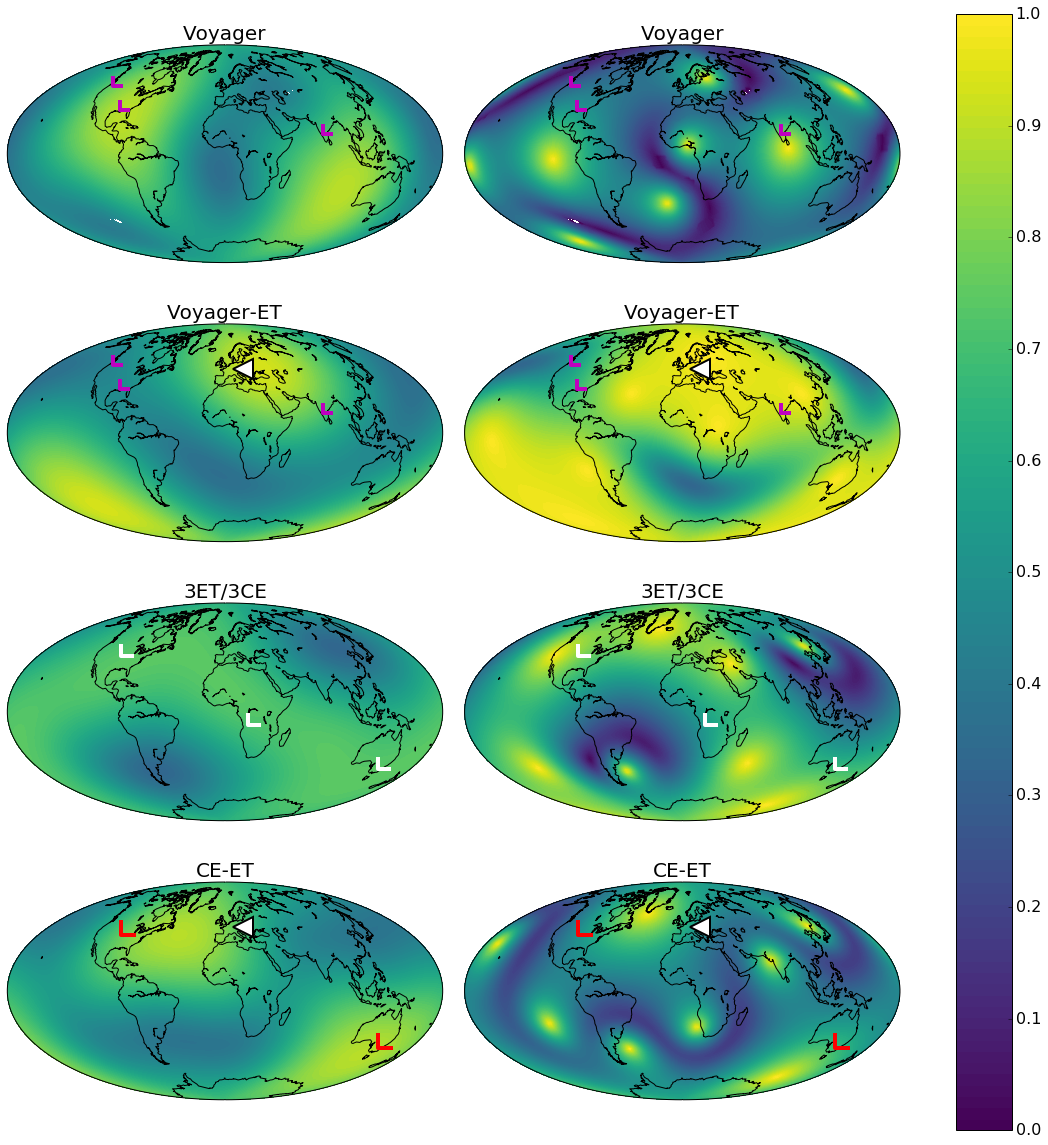}
\caption{Relative sensitivity of the different networks over the sky: Voyager, Voyager-ET, 3ET/3CE and CE-ET. 
\textit{Left:} network response as a function of sky position and
\textit{right:} alignment factor as a function of sky position.
Also shown are the locations of detectors in each network. Magenta markers are for Voyager detectors, white for ET, and red for CE.
}
\label{fig:alignment}
\end{figure*}

In Figure \ref{fig:alignment}, we plot the Network Response and Alignment Factor as a function of
sky location for the five networks under consideration.  
The Voyager network has the best sensitivity 
above and below the location of the two US LIGO detectors, as expected.  It has limited sensitivity to the second 
polarization over large parts of the sky, including the locations  with best network sensitivity.  
In the Voyager-ET network, the ET detector dominates the sensitivity so, as expected, 
we see the best sensitivity above and below the ET detector. The triangular ET is
equally sensitive to both polarizations, and so the Voyager-ET network has good sensitivity to the second polarization
over the majority of the sky.  Even in regions where ET has poor sensitivity, the second polarization is reasonably
well measured by a combination of ET and the three LIGO Voyager detectors. 

The 3ET and 3CE networks are comprised of detectors in identical 
locations, so the relative sensitivity over the sky will be identical for these networks. These networks 
have good coverage over much of the sky, but the peak sensitivity is noticeably lower than the other networks 
--- it is only 75\% of the maximum possible if all detectors were aligned, 
in comparison to over 90\% for the other networks.  This is to be expected, as the location of the detectors
has been chosen to maximize sky coverage; three co-located detectors would provide the greatest peak sensitivity 
but much worse sky coverage.  The homogeneous 3ET and 3CE networks have
markedly better sensitivity to the second polarization than the Voyager network.  This arises because the
detector orientations were optimized to give good sensitivity to the second polarization, whereas the LIGO Hanford and
Livingston detectors were deliberately aligned to be sensitive to the same polarization.  Finally, the heterogeneous CE-ET 
network shows best sensitivity
over the north atlantic and Australia, which is expected given the detectors are located in the US, Europe and Australia.  It
has relatively poor sensitivity to the second polarization over the sky.  However, in contrast to the Voyager network, CE-ET
has good sensitivity to both polarizations in areas of good overall sensitivity.

The Sky Coverage \cite{schutz_networks_2011} of a network is defined as the fraction of the sky for which 
the response is greater than $1/\sqrt{2}$ of the maximum. The Sky Coverage of the homogeneous ET and CE networks
is 79\%.  Even though the LIGO Voyager network also has three equal sensitivity detectors, the similar orientations of the LIGO
Hanford and Livingston detectors lead to a sky coverage of 42\%.  For the heterogeneous CE-ET and 
Voyager-ET networks, the sky coverage is 44\% and 37\% respectively.
This confirms what the plots suggest and indicates that the 3ET and 3CE networks have the most uniform response across the sky. 

\section{Source Localization}
\label{sec:loc}

To investigate the ability of different networks to localize sources, we use the formalism introduced in 
Refs.~\cite{Fairhurst:2009tc} and \cite{fairhurst_source_2011} and references therein.  In those papers, 
it was shown that localization is primarily determined by the timing accuracy, $\sigma_{t}$ in each detector which, in turn, 
is inversely proportional to the signal strength and frequency bandwidth $\sigma_{f}$ of the signal in the detector.  
Specifically, given a signal $h(t)$, the effective bandwidth is defined as
\begin{equation}
  \sigma_{f}^{2} = 
    \left(\frac{4}{\rho^{2}} \int_{0}^{\infty} df 
    \frac{f^2 |h(f)|^2}{S(f)}\right) - 
    \left(\frac{4}{\rho^{2}} \int_{0}^{\infty} df 
    \frac{f |h(f)|^2}{S(f)} \right)^{2},
\end{equation}
where the \ac{SNR}, $\rho$, in the absence of noise, is given by
\begin{equation}
    \rho^{2} = 4 \int_{0}^{\infty} \frac{| h(f) |^{2}}{S(f)} df \, .
    \nonumber
\end{equation}
The timing accuracy for a signal with SNR $\rho$ is then given by
\begin{equation}\label{eq:sigmat}
\sigma_{t} = \frac{1}{2\pi \rho \sigma_{f}} \, .
\end{equation}
Thus, $\sigma_{t} $ scales inversely with the SNR of the GW, $\rho$, and the effective bandwidth, $\sigma_f$, 
of the signal in the detector.

Using these expressions, it is possible to calculate the reduction in network SNR due to errors in sky location and
derive, at leading order, a relatively simple expression for the localization area.  The probability distribution for the location of the 
source (from a sky location $\mathbf{R}$) is given by
\begin{equation}\label{eq:loc}
p(\mathbf{r} | \mathbf{R}) \propto p(\mathbf{r})
 \exp \left[ - \frac{1}{2} (\mathbf{r} - \mathbf{R})^{T} \mathbf{M} (\mathbf{r} - \mathbf{R}) \right]
\end{equation}
where $\mathbf{r}$ is the reconstructed position of the source, $p(\mathbf{r})$ is the prior distribution (taken as uniform on the sphere), and the matrix $\mathbf{M}$ describes the localization accuracy and is
given by
\begin{equation}
\mathbf{M} = \frac{1}{\sum_{k} \sigma_{t_{k}}^{-2}} \sum_{i,j} 
\frac{(\mathbf{D}_{i} - \mathbf{D}_{j})(\mathbf{D}_{i} - \mathbf{D}_{j})^{T}}{2 \sigma_{t_{i}}^{2} \sigma_{t_{j}}^{2}}
\end{equation}
and $\mathbf{D}_{i}$ gives the location of the i-th detector.  Thus, the localization is improved by having greater separation between
the detectors and good timing accuracy, i.e. high \ac{SNR} and large bandwidth of the signal in the detectors.

Localization can be improved by accounting for the relative amplitude and phase of the signal observed in each detector.  These
are necessarily constrained by the fact that a gravitational wave has only two polarizations so, for networks of three or more
detectors, the relative amplitude and phase in the different detectors is constrained. 
When taken into account, this leads to a more rapid falloff in the network SNR away from the correct sky location
which, in turn, leads to an improvement in localization.  This has been discussed in detail in \cite{fairhurst:2017}, and a similar
analysis presented in \cite{wen_geometrical_2010}.  The resulting probability distribution for the localization
has the same form as eq.~(\ref{eq:loc}) with a modified expression for the matrix $\mathbf{M}$, which nonetheless
remains quadratic in the detector separations $\mathbf{D}_{i} - \mathbf{D}_{j}$.

Based on timing information alone, a source observed in three detectors can be localized to two regions in the sky.  The
two locations lie above and below the plane formed by the three detectors.  When we require the signal to be consistent
with two gravitational wave polarizations, this places restrictions on the relative amplitudes and phase differences between
the detectors.  In many cases, this information can be used to exclude the \textit{mirror} location and restrict the source
to a single sky position.  Of course, with four or more sites, timing information alone can be used to localize a source
to a single sky location.

In the following studies, we generate a population of events and determine which events
would be detected by a given network and how accurately they would be localized.  In all instances, we use the above formalism
and ignore the effects of noise which would change the recovered SNR and offset the optimal sky location from the expected values. 
We require that signals would be confidently detected by the detector network.  Specifically, we require
a network SNR of at least 12 as well as an SNR above 5 in at least two detectors in the network.%
\footnote{For a discussion of the effects of changing these thresholds and, in particular, removing the single detector thresholds
see \cite{macleod_fully-coherent_2016}.}
Furthermore, since the localization 
methods described above are accurate only to leading order, our localization results are based only upon detectors for which the signal 
has an SNR greater than 4.  As discussed in \cite{Fairhurst:2009tc}, at lower SNRs the approximations used here break down.  

The thresholds used mimic those used in the analysis of \ac{GW} data \cite{TheLIGOScientific:2016qqj}
to obtain events with a false alarm rate of less than 
1 per century \cite{Singer:2014qca, fairhurst_source_2011} and are the same as used in previous studies 
\cite{Fairhurst:2009tc, fairhurst_source_2011}.  In addition, they seem appropriate based on the initial gravitational wave
observations, where GW150914 and GW151226 both satisfied these requirements while LVT151012 had a network SNR of 10
and was not unambiguously identified as a signal \cite{TheLIGOScientific:2016pea}.  As \ac{GW} observations become more
common, and searches are further improved \cite{Nitz:2017svb}, it is possible that the detection thresholds will be reduced.  
While this will change the details of the results presented below, the \textit{relative} performance of the networks will remain 
similar.

\section{Results}
\label{sec:results}
\subsection{Face on Binary Neutron Star Mergers}
\label{ssec:faceon}

We first investigate the ability of the networks to localize a given source at a fixed distance, as a function of the sky 
location of a source. We simulate $1.4 - 1.4 \mathrm{M}_{\odot}$ \ac{BNS} systems that are oriented face on 
(i.e. with inclination, $\iota=0$) at a 
fixed distance at each point along a two dimensional 16 by 16 grid of sky coordinates.  We repeat the study for sources
at redshifts of $z=0.2$ ($D_{L} = 1 \mathrm{Gpc}$) and $z=0.5$ ($D_{L} = 3 \mathrm{Gpc}$).
At each sky location, we calculate the expected SNR in each of the detectors 
in the network.  For any signal that meets the detection and localization criteria given above, we calculate 
the 90\% localization region. Since the \ac{BNS} systems are face on, the GWs are circularly polarized, 
i.e. both polarizations have the 
same amplitude. Thus it is the overall sensitivity, and not the relative sensitivity to the two polarizations that
will affect the localization ellipses \cite{fairhurst:2014}.

\begin{figure*}[ht]
\centering
\includegraphics[width=0.49 \textwidth]{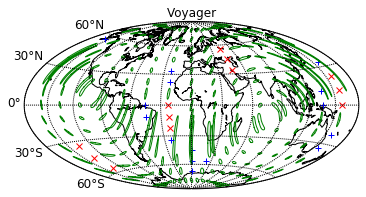}
\includegraphics[width=0.49 \textwidth]{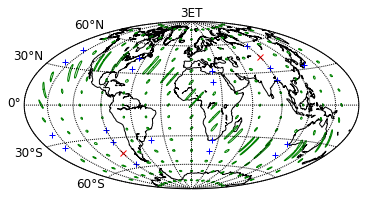}
\includegraphics[width=0.49 \textwidth]{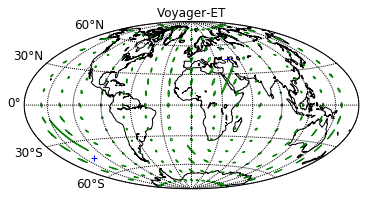}
\includegraphics[width=0.49 \textwidth]{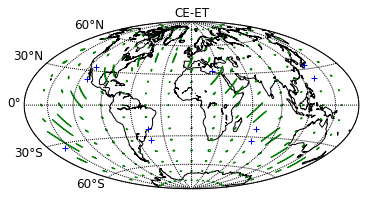}
\includegraphics[width=0.49 \textwidth]{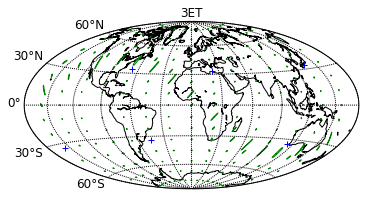}
\includegraphics[width=0.49 \textwidth]{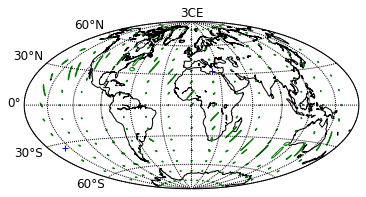}
\caption{The localization ellipses at different sky locations for face-on 1.4-1.4 BNS binaries at a redshift of: \textit{left} -  z = 0.2 (luminosity distance of 1 Gpc) and \textit{right} - z = 0.5 (luminosity distance of 3 Gpc). The red crosses indicate that the BNS at this sky position 
would not be detected --- either due to a network SNR less than 12, or not having SNR $> 5$ in at least two detectors.
The blue + symbols indicated sources that would be detected, but not well localized due to being identified in only
two detectors. The ellipses give the $90\%$ localization regions for a source from a given sky location.}

\label{fig:ellipses}
\end{figure*}

Figure  \ref{fig:ellipses} shows the localization regions for these \ac{BNS} sources in the five networks under
consideration.  In the figures, a red cross indicates that the detection criteria (network SNR $>12$ and two detectors with SNR $>5$) 
were not met for a BNS at this sky position and redshift; a blue plus indicates that the source would be detected 
but fails our localization criterion (SNR $>4$ in three or more sites). For signals which would be confidently 
detected, and observed in at least
three sites, the green ellipses show the $90\%$ confidence region for the localization.

For \ac{BNS} mergers at $z=0.2$, the LIGO Voyager network would observe the signal over the majority of the sky.
There are, however, four patches where the signal would not be found, which correspond
to areas of poor sensitivity for the two US LIGO detectors.  Furthermore, there are regions where the signals would be detected 
but not localized, based on our conditions, and these correspond to locations where LIGO India has poor sensitivity.   
For those signals which are localized, the areas are typically large, as these events will be close to the 
detection threshold in the network.  We can clearly identify a band for which the localizations are extended
in one dimension.  These points are close to the plane defined by the three detector locations. A large change in sky 
location, in a direction perpendicular to the plane of the detectors, leads to a relatively small change in the relative 
arrival times and consequently poor localization.  These results are consistent with those obtained for the advanced LIGO
network (incorporating LIGO India) given in \cite{fairhurst:2014}.

The Voyager-ET network is able to detect sources at $z=0.2$ over the essentially the whole sky.  For localization,
we require the signal to be observed at three sites; although all three of the detectors in the triangular ET will 
observe the signal, they provide rather poor localization by themselves.  Thus, the network is limited by the 
requirement that two LIGO Voyager detectors observe the signal.  The sky locations where sources are
not localized correspond to the locations for which the US LIGO detectors have poor sensitivity, and
these sources are only detected in ET and LIGO India.  
The 3ET network also gives excellent coverage over essentially the whole sky.  
There are still a handful of points for which localization is not possible.  Again, these correspond to points where one 
of the detectors has close to zero sensitivity.  As before, we see the characteristic extended ellipses at locations which lie
close to the plane defined by the three ET detectors. 

For signals at $z=0.5$ we consider the three networks comprised of ET and CE detectors.  In all cases, the
sources are observed over essentially the whole sky.  For the 3ET network, there are significant regions where the
source is not well localized as it is seen in only two detectors, but the size of these regions shrinks for the CE-ET and 3CE
networks due to the increased sensitivity of the CE detector.  Finally, as expected, the signals are relatively
poorly localized in directions close to the plane defined by the three detectors.  

For a two-site observation, the localization is typically restricted to a fraction of a
ring in the sky with an area of hundreds of square degrees \cite{Abbott:2016blz, TheLIGOScientific:2016pea} and we consider these
sources to not be localized.  The degeneracy along the ring
is broken by relative amplitude and phase measurements in the different detectors.  For events observed
with the triangular ET detector and a single L-shaped detector, the localization may be greatly improved ---
the triangular detector recovers the amplitude and phase of both \ac{GW} polarizations so a single, additional
observation will provide enough information to break the sky location degeneracy.  Furthermore, when there
are additional detectors in the network that did not observe the event, this information can be used to further improve
the localization.  We do not consider these effects here, but note that it would be interesting to examine in 
detail localization with a network comprised of one triangular and one L-shaped detector.

In these plots we are ignoring the fact that sources detected at three sites may be localized to two distinct patches 
in the sky, one above and one below the plane formed by the three sites.  In many cases, the degeneracy
can be broken based on consistency of the observed amplitude and phase of the signal in each of the detectors.
For the systems at $z = 0.2$, both the 3ET and Voyager-ET networks will provide localization to a
single region for essentially all sky locations. Voyager localizes to one patch on the sky 70\% of the time, 
and so about a third of the localization ellipses shown below will be augmented by a similar sized region in 
the \textit{mirror} location. At z=0.5, the CE and ET networks all localize to a single patch for at least 95\% of sky locations. 

\subsection{A Population of Coalescing BNSs}
\label{ssec:population}

\begin{figure*}[ht]
\centering
\includegraphics[width=.49 \textwidth]{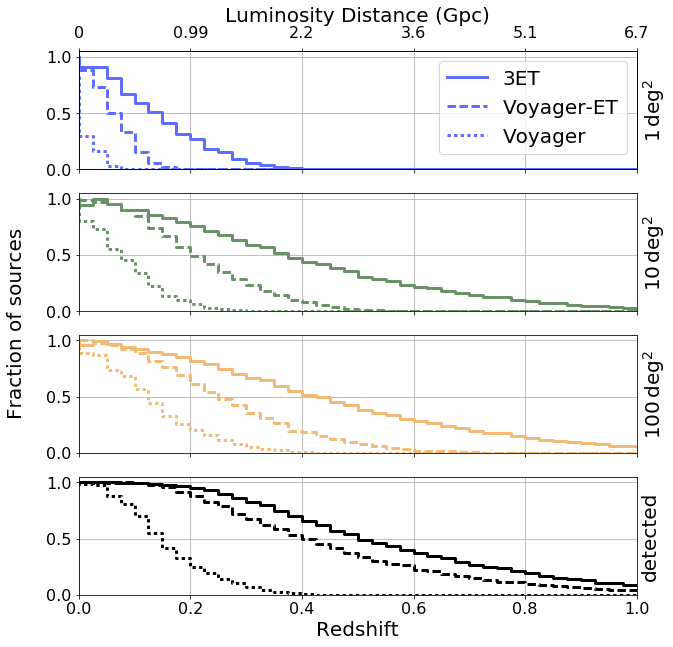}
\includegraphics[width=.49 \textwidth]{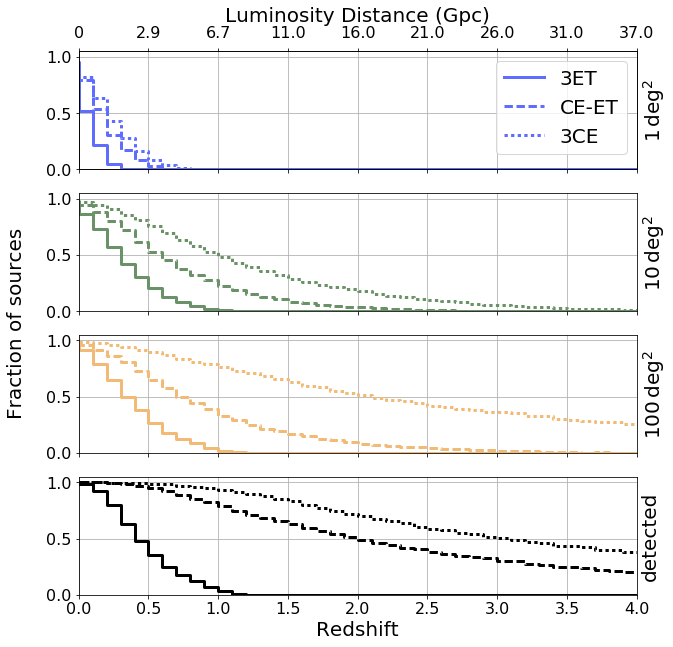}
\caption{The detection and localization efficiency as a function of redshift and luminosity distance of the Voyager, Voyager-ET and 3ET networks (left column) and the 3ET, CE-ET and 3CE networks (right column). For visual comparison, 3ET is plotted as a solid line in both. \textit{Right column}: Voyager-ET and Voyager are the dashed and dotted lines respectively. \textit{Left column}: CE-ET and 3CE are the dashed and dotted lines respectively. From bottom to top, the rows show the fraction of 
events at a given redshift/distance that will be detected and localized within 100, 10 and 1 $\deg^{2}$.}
\label{fig:efficiency}
\end{figure*}

Now, let us consider network localization for a population of \ac{BNS} coalescences. We follow Singer 
et.~al.~\cite{Singer:2014qca} in choosing the \ac{BNS} component masses uniformly in the astrophysically 
motivated range $1.2 - 1.6 M_{\odot}$. This encompasses the masses of all observed neutron stars in binaries
and the 1-sigma interval of the initial mass function for a variety of formation mechanisms \cite{ozel_mass_2012,pejcha_observed_2012}. 
The orientation of the sources is uniformly distributed: uniform in polarization, cosine of source inclination and the phase of the 
\ac{GW} at merger. 
We distribute the sources isotropically and uniformly in comoving volume with a uniform \textit{local} merger rate, 
meaning that the observed rate at a given redshift will be decreased by a factor of $(1+z)$. The networks are affected by the duty cycles of individual detectors. However, we do not consider this effect here and assume detectors to have full duty cycles. 

Figure \ref{fig:efficiency} shows the detection efficiency --- the fraction of events that would be observed --- for each network as a 
function of redshift or distance.  For those \ac{BNS} mergers which are detected by a given network, we calculate the 90\% 
confidence sky localization using the prescription given in Section \ref{sec:loc}.  We also show the fraction of events that would be 
localized within $1, 10$ and $100 \deg^{2}$ for each 
network as a function of redshift.  

On the left hand side of Figure \ref{fig:efficiency}, we consider the Voyager, Voyager-ET and 3ET networks.  
The Voyager-ET and 3ET networks have rather comparable 
sensitivities, both networks identify over $90\%$ of sources within a redshift of $z = 0.2$ and the majority of signals within 
a redshift of $z = 0.4$.  The LIGO Voyager network has good all sky sensitivity within a redshift of $z=0.1$, after which it drops
rapidly with essentially no sensitivity beyond $z=0.4$.   Since we require a source to be
observed in three sites for good localization, it is unsurprising that the Voyager and 3ET networks are capable of
localizing the majority of observed sources --- in particular, essentially all sources are localized within $100 \deg^{2}$ and the
majority within $10 \deg^{2}$.  For the heterogeneous Voyager-ET network, the fraction of sources localized is much lower than
the fraction detected.  For example, at $z = 0.4$ over half of all sources are detected but only 10\% are localized within $10 \deg^{2}$.
These are the events which are too distant to be observed by the LIGO Voyager detectors so, while they
are observed by ET they cannot be localized.   For all three networks, only a fraction of events will be localized to within $1 \deg^{2}$ 
and those will be primarily nearby, loud events.  For a 3ET network, half of the events at a redshift of $z = 0.15$ with be localized to within 
$1 \deg^{2}$.

The right panel of Figure \ref{fig:efficiency} shows the same results for the CE and ET based networks.  The results are
comparable to those described above: detection efficiency is limited by the second most sensitive 
detector in the network, while localization requires a third detector to observe
the signal.  In particular, we note that while the CE-ET and 3CE networks have similar overall detection efficiencies, the 3CE
network provides much better localizations.  For example, 3CE localizes 50\% of sources at $z = 2$ to $100\deg^{2}$ while 
the CE-ET network is unable to give good localizations for signals at this redshift.  We note that
those signals which are localized in the heterogeneous CE-ET network are typically localized within $10 \deg^{2}$ as they will be
recovered with high \ac{SNR} in the Cosmic Explorer detectors.  Finally, it is again only the 
loudest, nearby signals which are localized within $1 \deg^{2}$.  The CE-ET and 3CE networks localize half of signals within
$1 \deg^{2}$ to a redshift of $z \sim 0.25$.

\begin{figure*}[ht]
\centering
\includegraphics[width=.49 \textwidth]{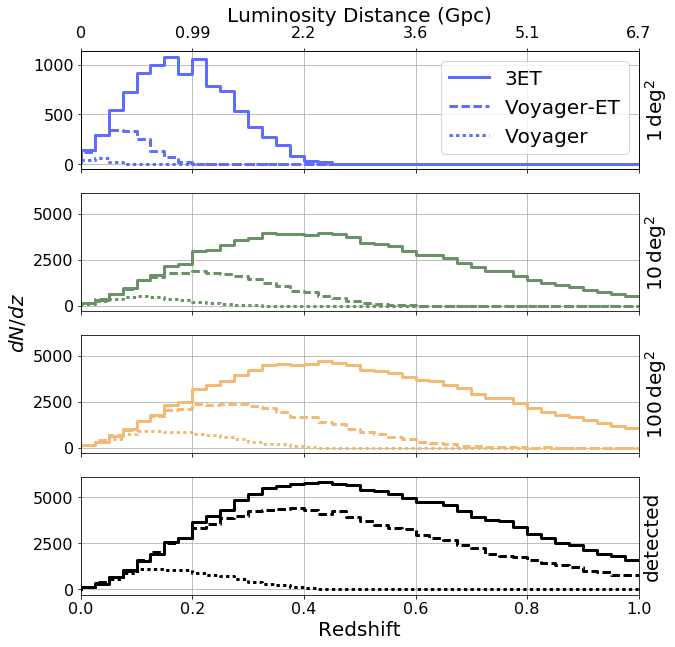}
\includegraphics[width=.49 \textwidth]{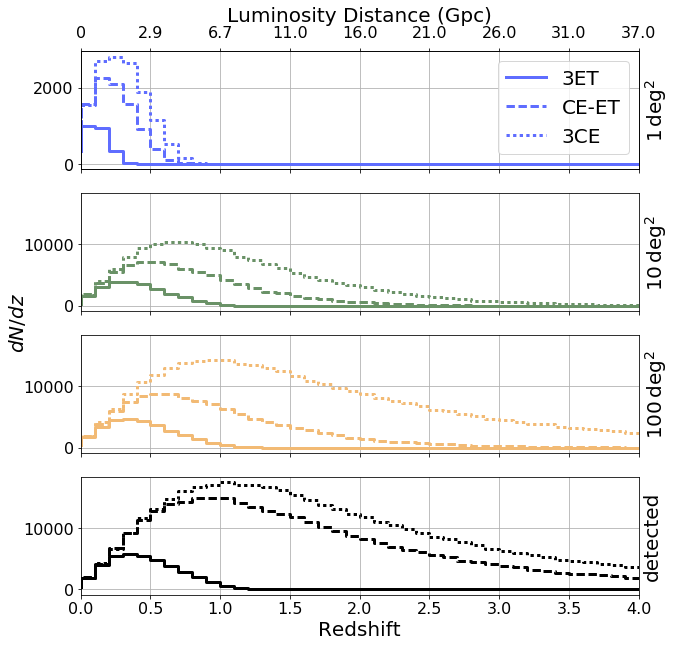}
\caption{
The number of BNS observations and localizations per year with future networks, as a function of redshift and luminosity distance. The y-axis is scaled so that the
area under the curves gives the number of events per year. This assumes a source distribution uniform in comoving volume with an intrinsic rate of 1 merger per $10^{7} \mathrm{Mpc}^{3} \mathrm{y}$. Note that the y-axis on $1 \deg^{2}$ plot is different from the others. For visual comparison 3ET is plotted as a solid line in both column. \textit{Right column}: Voyager-ET and Voyager are the dashed and dotted lines respectively. \textit{Left column}: CE-ET and 3CE are the dashed and dotted lines respectively. From bottom to top, the rows show the number density of 
events at a given redshift/distance that will be detected and localized within 100, 10 and 1 $\deg^{2}$.}
\label{fig:population}
\end{figure*}

Figure \ref{fig:population} shows the expected number of observed events as a function of redshift 
for the five networks, and the overall results are summarized in Table \ref{tab:bns_pop}.
In order to obtain these results, we have taken a population of sources
with a uniform local merger rate density.  This is appropriate on the scale of hundreds of Mpc where the local variations 
become insignificant \cite{LIGOS3S4Galaxies}, but less so when considering sources within the range of the ET and CE detectors.  
The \ac{BNS} formation rate, and consequently the merger rate, is expected to follow the star formation rate 
\cite{ligo_scientific_collaboration_predictions_2010} which varies with redshift \cite{heavens_star-formation_2004},
peaking near a redshift of one.  However, due to the large uncertainty in the rates, their dependence on the star 
formation rate and the delay between formation and merger, we have chosen not to incorporate these effects. 
Furthermore, we use an intrinsic rate of $10^{-7} \mathrm{Mpc}^{-3} \mathrm{yr}^{-1}$, which lies within
the current range of predicted rates \cite{Abbott:2016ymx}, and we use this fiducial
rate when comparing the network sensitivities.  There is at least an order of magnitude uncertainty on the rate of
\ac{BNS} mergers. Overall changes to the merger rate will simply scale the number of 
observations for all networks equally but changes in the redshift evolution of the rate will affect relative 
performances.

As expected, the number of events detected by the 3ET and Voyager-ET networks are comparable, with the
3ET network is sensitive to $40\%$ more \ac{BNS} mergers than Voyager-ET, while the Voyager network
observes around one tenth as many mergers.  However, due to the differences in localization, the Voyager-ET
localizes only a third as many events as the 3ET network and, for events localized within $10 \deg^{2}$ the
peak of the redshift distribution is at 0.2 rather than 0.4 (and within $1 \deg^{2}$ the peak is at 0.06 rather than 0.2).
We see similar results for the CE-ET and 3CE networks: they are both able to detect a comparable number of events, but 
significantly fewer are localized by the heterogeneous network.

\begin{table}
\begin{tabular}{|c|c|c|c|c|c|}
\hline
Network & Voyager & Voy-ET & 3ET & CE-ET & 3CE \\
\hline
Detected & 260 & 2700& 3800 & 34000& 47000\\
Localized & 240& 910& 3100& 12000& 40000\\
Within 1 $\deg^2$ & 4 & 40& 240 & 920& 1400 \\
Within 10 $\deg^2$ & 87 & 560& 2300 & 7800& 16000\\
Within 100 $\deg^2$ & 210 & 900& 3000 & 11000& 36000\\
Median Area ($\deg^2$) & 20& -& 6 & - & 17\\ 
Single Patch & 46\%& 99\%& 85\%& 97\%& 89\%\\ 

\hline
\end{tabular}
\caption{\label{tab:bns_pop}  Performance metrics for Voyager, Voyager-ET, 3ET, CE-ET and 3CE networks for a population
of \ac{BNS} coalescences distributed uniformly in comoving volume with an intrinsic rate of 1 merger per $10^{7} \mathrm{Mpc}^{3} 
\mathrm{y}$. 
\textit{From top to bottom}: The number of sources per year that are detected and localized by each network; 
   the number of sources localized per year within 1, 10 and 100 $\deg^{2}$ respectively;
    the median localization area of all detected sources and the fraction of localized sources whose position is restricted
    to a single patch in the sky. Note, the median source is not localized by the two heterogeneous networks CE-ET and Voyager-ET.}
\label{table:network_performance}
\end{table}

\subsection{Implication for \ac{EM} Followup}
\label{sec:interpret}

The primary motivation for accurate localization of \ac{GW} signals is to facilitate the observation of
electromagnetic counterparts.  The requirements on localization will depend upon
the strength of the electromagnetic emission accompanying a \ac{BNS} merger, as well as the ability of wide-field
telescopes to cover the error region.  The most likely counterparts from \ac{BNS} mergers 
\cite{Metzger:2011bv, Fernandez:2015use} are short \ac{GRB} and kilonova emissions.  While short
GRBs can be observed to cosmological distances, they are believed to be rather tightly beamed, so that
only a small fraction of \ac{BNS} mergers would be accompanied by a GRB counterpart \cite{Clark:2014jpa}.
However, since the \ac{GRB} emission is likely to be essentially concurrent with the merger, it will be 
difficult to use \ac{GW} observations to provide advanced warning to \ac{GRB} satellites.
Thus, it seems likely that the \ac{GW} and GRB signals will be independently observed, and the 
better localization will typically come from the \ac{GRB} signal.  Thus, although joint \ac{GW} and \ac{GRB} searches have been
performed in the past for finding any associated GW signal with observed GRBs 
have been conducted in the past \cite{Abbott:2016cjt}, and will continue to be interesting enterprise for multi-messenger astronomy, 
they will not be significantly influenced by \ac{GW} localization.  Consequently, we will focus the remainder of our discussion on 
kilonova observations.  

The neutron-rich ejecta from \ac{BNS} and \ac{NSBH} mergers will undergo r-process nucleosynthesis, producing 
heavy elements which will subsequently decay;  this decay process will power an electromagnetic transient
known as a kilonova (see e.g.~\cite{Metzger:2016pju} for details).  There are various models for
the kilonova emission, which depend upon the mass of the ejecta as well as its opacity \cite{Fontes:2017zfb, Barnes:2016umi}.  
Broadly, the prediction
is for an optical or near infrared emission, which will last for days or possibly weeks.  The luminosity of the kilonova
emission is uncertain, but we take a fiducial model with magnitude 22 emission from a source at $200 \mathrm{Mpc}$ \cite{Metzger:2016pju}.  
To date, there has been one putative near-infrared kilonova observation from GRB 
130603B which was observed with a magnitude of $25.8$ at redshift of $z \approx 0.35$ \cite{Tanvir:2013pia, Berger:2013wna}, 
which is broadly consistent with this picture.

Although GW151226 was a BBH merger, significant electromagnetic followup was performed.  Notably, DeCAM, J-GEM and
PanSTARRS searched for optical and near infrared counterparts to the event \cite{Yoshida:2016ddu, Cowperthwaite:2016shk, 
Smartt:2016aa}.
While no counterpart was found, these searches were able to cover tens of square degrees at the peak of the \ac{GW}
localization error region and place upper limits on $i$ and $z$ band emission at magnitudes between 20 and 24.

Taking our fiducial kilonova model, the current generation of wide-field telescopes, such as Pan Starrs, zPTF, SkyMapper, 
Black Gem, which have limiting magnitudes around 24 would be able to observe kilonova emission to $z \approx 0.1$ or a
luminosity distance of $500 \mathrm{Mpc}$.  
The results in Figures \ref{fig:efficiency} and \ref{fig:population} show that the Voyager network has good sensitivity within
the range of the current generation of telescopes, and would identify and localize the majority of \ac{BNS} mergers
at $z \lesssim 0.1$ to within $10 \deg^{2}$.  All of the other networks are able to detect and localize within $10\deg^{2}$ essentially every 
event at $z \lesssim 0.1$, thereby enabling followup with one, or a handful, of pointings.  
For LSST, with a limiting magnitude around 26, kilonovae could be observed to $z \approx 0.25$ or $1.3 \mathrm{Gpc}$.
At these distances, the sensitivity of the Voyager network is insufficient to identify the majority of signals, let alone provide
accurate localizations.  
The Voyager-ET network would observe the majority of \ac{BNS} within this range although, for the more distant signals, the
limited sensitivity of the Voyager detectors means that many would be observed by only the ET detector and consequently be 
very poorly localized.  The networks with three ET or CE detectors provide excellent sensitivity to $z = 0.25$ and are capable
of localizing the vast majority of sources to within $10 \deg^{2}$. 

Of course, the details of kilonova emission are still uncertain and there
are models that predict significantly stronger or weaker emission.  For example, there are models where the emission is powered by 
fallback accretion \cite{Metzger:2016pju} that predict magnitudes of 21 (for \ac{BNS}) or 20 (for \ac{NSBH}) at 200 Mpc.  
This increases the distance at which kilonovae could be observed by a factor 
of $1.5$ and $2.5$ respectively making kilonovae associated to \ac{BNS} observable to 750 Mpc (or 2 Gpc with LSST).  A 3ET network 
would identify and localize three times as many kilonova observations within the LSST range as from the Voyager-ET network.  
This serves to highlight the point that the case for localization capacity of future \ac{GW} networks is intimately tied to
our knowledge of the \ac{EM} emission from these mergers.

As some of the strongest emissions are predicted from \ac{NSBH} mergers, it is interesting to briefly consider them.  While
we have not performed simulations with \ac{NSBH} systems, it is straightforward to provide approximate sensitivities based
on the \ac{BNS} results given above. The sensitivity
of gravitational wave detectors scales, at leading order, as $\mathcal{M}^{5/6}$, where $\mathcal{M}$ is the chirp mass.  
Consequently, for a signal at a fixed distance, orientation and sky location, the \ac{SNR} with which \ac{NSBH} will be observed
can be approximated as
\begin{equation}
  \rho_{\mathrm{NSBH}} \approx 1.1 \left(\frac{M_{\mathrm{BH}}}{M_{\mathrm{NS}}}\right)^{1/3} 
  \left(1 + \frac{M_{\mathrm{BH}}}{M_{\mathrm{NS}}}\right)^{1/6} \rho_{\mathrm{BNS}} 
\end{equation}
where $M_{\mathrm{BH}}$ is the black hole mass and $M_{\mathrm{NS}}$ is the neutron star mass.  Thus, the observed \ac{SNR}
for a \ac{NSBH} with $M_{\mathrm{BH}} = 5 M_{\odot}$ is $2.2$ times that of a \ac{BNS}, and $3.0$ times for 
$M_{\mathrm{BH}} = 10 M_{\odot}$.  Consequently, to a reasonable approximation, we can scale the distances in Figure 
\ref{fig:efficiency} by between 2 and 3 to obtain \ac{NSBH} sensitivities.  Thus, for the most optimistic \ac{NSBH} 
kilonova emissions, a network with three ET or CE detectors would identify and localize most sources within the LSST range; 
Voyager augmented by ET would identify but not localize the more distant sources and Voyager alone would have a range 
comparable to existing wide-field telescopes.

\section{Discussion}
\label{sec:discussion}

We have compared the sensitivity of proposed future gravitational wave networks to \ac{BNS} signals, and their ability to
accurately localize events.  We find that a minimum of two detectors, which includes the triangular ET, at an improved 
sensitivity are sufficient to provide a substantial increase in the number of observed sources.  For example, the addition of ET to a
network of LIGO-Voyager sensitivity detectors could increase the rate of observations by an order of magnitude.  
However, in order to obtain good source localization, we require a minimum of three sites to observe the event.
Consequently, in networks with one or two detectors that are significantly more sensitive than the others, we find
that the majority of detected sources are not well localized.  In contrast, when the three most sensitive detectors in
the network have comparably sensitivity, the majority of signals are well localized with a median localization
area around $10 \deg^{2}$.  

Previously it has been argued that building more detectors further apart improves localization (see for example 
\cite{klimenko_localization_2011, fairhurst_source_2011}). However, we find that this is only true when the sensitivities 
of the detectors in the network are approximately homogeneous, as is often assumed for the advanced detector (2nd generation)
networks \cite{Fairhurst:2009tc,fairhurst_source_2011,fairhurst:2014}. In the case of heterogeneous sensitivities, 
such as the Voyager-ET network, the localization will often be limited since the event cannot be detected 
by the less sensitive detectors. In such a network, we expect that for a majority of events we will
obtain limited directional information. 

The interpretation of our results depends critically upon the science 
question of interest.  In particular, the utility of accurate \ac{GW} localization as a function of redshift will depend critically 
upon the strength and spectrum of the associated \ac{EM} emission, and the sensitivity and field of view of the 
associated telescopes and satellites.  For a standard kilonova model, the LIGO Voyager network provides adequate sensitivity to 
identify and localize potential kilonova signals for currently operating telescopes but the network must be augmented by at least one 
ET or CE detector to provide adequate sensitivity to localize all kilonovae that could be observed by LSST.  For the models predicting the 
strongest kilonova emission, a three detector network of ET or CE detectors could increase, by a factor of a few, the number of events 
observed jointly with \ac{GW} signals.

In this study, we have neglected a number of factors that effect the size of localization errors in real detector networks. 
Though the time of arrival and amplitude and phase of GWs carry most of the information relevant to localization, other information 
can reduce the size of the localization errors. These include, realistic prior distributions on other astrophysical parameters ---
particularly source inclination and distance, correlations with other parameters such as component masses 
\cite{grover_comparison_2014}, spin and precession effects. Furthermore, we continue to assume that a signal must be identified 
in two detectors to be detected, and three to be localized.  Ideally, performing a fully coherent analysis of the data \cite{macleod_fully-coherent_2016, DalCanton:thesis}, 
would improve the performance of heterogeneous networks when the SNR in the less sensitive detectors is low and would, in
effect, remove our requirement of a signal being clearly identified in at least two detectors.  While it is possible to localize
sources with only two detectors, the first \ac{GW} observations make it clear that the localization areas will typically be hundreds
of square degrees, so our approximation that these sources are not localized is reasonable.

Furthermore, we have neglected systematic uncertainties. Errors introduced by mismatches between template waveforms and signals 
\cite{Fairhurst:2009tc} are expected to introduce a similar effect in all detectors and therefore the effect on the time difference, 
and localization, is likely to be negligible. On the other hand, errors in the calibration of \ac{GW} detectors \cite{Abbott:2016jsd}
will be uncorrelated, 
and these errors can significantly impact localization. For instance, roughly one third of the localization error budget for GW150914 
was due to strain calibration uncertainty \cite{the_ligo_scientific_collaboration_binary_2016, Abbott:2016jsd}.  
At high SNR calibration errors are 
expected to dominate the overall error budget for localization \cite{Fairhurst:2009tc}. Thus, the ability to achieve the reported sub 
square-degree localizations predicted here will depend critically on the calibration accuracy of the detectors, with likely requirements
of uncertainties under $1\%$ in amplitude and $1^{\circ}$ in phase.  The impact of calibration on gravitational wave localization with
future networks deserves further study.

Finally, in this paper we have restricted attention to localization of \ac{BNS} signals.
For many science questions, accurate localization itself is not critical but is required for 
accurate measurement of the distance to the source.  The gravitational 
wave signal from the inspiral and merger of \ac{BNS}, \ac{NSBH} and \ac{BBH} leads to the accurate measurement of only the 
luminosity distance $D_{L}$ and the redshifted masses $M (1 + z)$.  One goal of gravitational wave astronomy is to map the
merger history of black holes and neutron stars through cosmic time.  Accurate distance measurements are required not only
to infer the redshift of the source, but also to obtain the mass of the source.  
For tests of cosmology, we require an independent measurement of the redshift.  
There are numerous methods proposed for this measurement, including
identification of a host galaxy from EM counterpart; statistical association with a host galaxy \cite{DelPozzo:2011yh}; assumption of a
narrow mass range of neutron stars in binaries \cite{Taylor:2012db}; observation of post-merger features in the waveform 
\cite{Messenger:2011gi}.  In all cases, an accurate measurement of the distance (and consequently good source localization) is
essential.   A detailed investigation of these issues is beyond the scope of this paper.  Nonetheless, it seems likely that a network of three or more
detectors of comparable sensitivity will increase scientific returns from a future gravitational wave network.

\section*{Acknowledgements}

The authors would like to thank Rana Adhikari, Lisa Barsotti, Yanbei Chen, Matt Evans, Laura Nuttall, Bangalore Sathyaprakash, 
Patrick Sutton and Salvatore Vitale for useful discussions. Also the authors are grateful to Tito Dal Canton for many helpful suggestions to improve the presentation of the paper. This work was supported by the STFC grant ST/L000962/1.

\appendix

\section{Optimal Detector Orientations}
\label{sec:appendix}

Table \ref{table:locations} contains all the final locations and orientations of all detectors used in this study. In order to fix the orientations of the detectors for the ET and CE networks, we make use of the following FoM, taken from Ref. \cite{raffai_optimal_2013},
\begin{equation}\label{eq:orientation}
I = \bigg( \frac{1}{4\pi}    \oint | f_{+}^{net} - f_{\times}^{net} |^2 d\Omega \bigg) ^{-1/2}
\end{equation}
Holding the USA detector fixed at 0$^\circ$ we rotated the Central Africa and Australia detectors from 0-90$^\circ$ (due to rotational symmetry of the polarizations all other rotations map to this basis). We optimize the ability of the ET and CE networks to observe both polarizations by choosing the orientation angles to maximize \ref{eq:orientation}. 

\begin{table}
\begin{tabular}{|c|c|c|c|}
\hline
Detector& Longitude& Latitude & Orientation \\ 
\hline
LIGO Livingston Voyager & -90.77 & 30.56 & 162.3 \\ 
LIGO Hanford Voyager& -119.41 & 46.46 & 234.0 \\ 
LIGO India Voyager & 76.4 & 14.2 & 346.8 \\ 
European $\Delta$ ET& 18.7 & 48.5 & - \\ 
Central Africa CE/ET& 17.2& -9.9 & 82.4 \\ 
Australia CE/ET& 146.9 & -35.8 & 84.4 \\ 
USA CE/ET& -98.4 & 38.9 & 0 \\ 
\hline
\end{tabular}
\caption{\label{table:locations}The locations and orientations of the detectors used in this study. 
All numbers are given in degrees and the orientation angle is defined clockwise relative to a 
hypothetical L-shaped interferometer with arms due North and East.}
\end{table}

\section*{References} 
\bibliographystyle{unsrt_edit}
\bibliography{iulpapers,refs,MyLibrary}

\end{document}